\documentclass[a4paper,conference]{IEEEtran}

\IEEEoverridecommandlockouts
 
\usepackage[utf8]{inputenc}
\usepackage[T1]{fontenc}
\usepackage{url}
\usepackage{tikz}
\usepackage{pgfplots}
\usepgfplotslibrary{groupplots} 
\usepgfplotslibrary{dateplot} 
\pgfplotsset{compat=newest}
\usepackage{pgf} 
\usepackage{ifthen} 
\usepackage{cite}
\usepackage[cmex10]{amsmath} 
\usepackage{amsfonts, amssymb} 
\usepackage[version=4]{mhchem}
\usepackage{graphicx, amsthm, float, dsfont,color}

\newtheorem{theorem}{Theorem} 
\newtheorem{lemma}{Lemma}

\makeatletter
\newcommand{\vast}{\bBigg@{3}}
\newcommand{\Vast}{\bBigg@{4}}
\makeatother

\newcommand{\bs}[1]{\boldsymbol{#1}} 
\newcommand{\mcl}[1]{\mathcal{#1}}

\newcommand{\I}{\mathrm{I}}
\newcommand{\Q}{\mcl{Q}}

\begin{document}

\title{Random Gilbert-Varshamov Codes\\ for Joint Source-Channel Coding}

\author{%
  \IEEEauthorblockN{Seyed AmirPouya Moeini}
  \IEEEauthorblockA{University of Cambridge\\
    \texttt{sam297@cam.ac.uk}}
  \and
  \IEEEauthorblockN{Albert Guill\'en i F\`abregas}
  \IEEEauthorblockA{University of Cambridge\\
  Universitat Polit\`ecnica de Catalunya\\
    \texttt{guillen@ieee.org}}
  \thanks{This work was supported in part by the European Research Council under Grant 101142747 and in part by the Spanish Government under Grant PID2020-116683GB-C22.}
  }

\maketitle

\begin{abstract}
	We propose a random coding technique for joint source-channel coding of discrete memoryless sources and channels.
	The approach builds on the random Gilbert-Varshamov code construction of Somekh-Baruch \emph{et al.} and extends it to the joint source-channel setting.
	We show that the resulting ensemble attains the maximum of the random-coding and expurgated error exponents.
\end{abstract}

\section{Introduction}
We study joint source-channel coding (JSCC) for the transmission of non-equiprobable messages generated from a discrete memoryless source with distribution
 $P^k(\bs{v}) = \prod_{i=1}^k P_V(v_i)$, 
where $\bs{v}=(v_1,\ldots,v_k)\in\mcl{V}^k$ denotes the source message and $\mcl{V}$ is a finite alphabet.
The channel is a discrete memoryless channel characterized by $W^n(\bs{y}|\bs{x}) = \prod_{i=1}^n W(y_i | x_i)$, where $\bs{x}=(x_1, \ldots, x_n) \in \mcl{X}^n$ and $\bs{y}=(y_1, \ldots, y_n) \in \mcl{Y}^n$ are the channel input and output sequences, respectively,
and $\mcl{X}$ and $\mcl{Y}$ are discrete alphabets.

In JSCC, an encoder maps each length-$k$ source message $\bs{v}$ to a length-$n$ codeword $\bs{x}_{\bs{v}}$, which is then transmitted over the channel. 
Based on the channel output $\bs{y}$ the decoder estimates which source message was transmitted.
We refer to $t \triangleq k / n$ as the transmission rate.
An error exponent $E>0$ is said to be \emph{achievable} if there exists a sequence of codes of blocklength $n$ such that the error probability satisfies
\begin{align}
	p_e \leq e^{-n E+o(n)},
\end{align}
where $o(n)$ satisfies $\lim _{n \rightarrow \infty} o(n) / n=0$.

Two classical achievable error exponents in channel coding are the random-coding and the expurgated exponents.
In \cite{1056281}, Csiszár and Körner proved that there exists a code that attains the maximum of these two exponents.  
More recently, \cite{8656563} introduced a random-coding construction in which codewords are generated sequentially from a fixed type class, 
subject to a minimum-distance constraint with respect to previously generated codewords, and showed that the resulting ensemble achieves an exponent that is no smaller than that of \cite{1056281}.

The study of error exponents for JSCC was initiated by Gallager \cite[Prob.~5.16]{gallager} and Csiszár \cite{csiszar1980}. In \cite{1056585}, Csiszár introduced two achievable expurgated error exponents, but was unable to compare them. In \cite{izs26}, we compared the two exponents and showed that one is a particular case of the other. In this work we refer to the expurgated exponent as the most general of Csiszár's two error exponents.
To the best of our knowledge, no JSCC random coding construction has previously been shown to attain the maximum of the random-coding and expurgated exponents.  
In this paper, building on the idea of \cite{8656563}, we extend the recursive distance-based construction to the JSCC setting.  
We show that the resulting ensemble simultaneously achieves both the random-coding and expurgated error exponents.

\subsection{Notation}
Scalar random variables are denoted by uppercase letters, their realizations by lowercase letters, and their alphabets by calligraphic letters. Random vectors are written in boldface. We use the notation $[a]$ to denote all integers from $0$ to $a$.
For two positive sequences $\{f_n\}$ and $\{g_n\}$, we write $f_n \doteq g_n$ if $\lim _{n \rightarrow \infty} \frac{1}{n} \log \frac{f_n}{g_n}=0$, 
and we write $f_n \, \dot{\leq}\,\, g_n$ if $\lim \sup _{n \rightarrow \infty} \frac{1}{n} \log \frac{f_n}{g_n} \leq 0$.
The type of a sequence $\bs{x}=\left(x_1, ..., x_n\right) \in \mcl{X}^n$ is its empirical distribution, defined by
$\hat{P}_{\bs{x}}(x) \triangleq \frac{1}{n} \sum_{i=1}^n \mathds{1}\left\{{x}_i=x\right\} .$
 The set of types for vectors in $\mcl{X}^n$ ie denoted by $\mcl{P}_n(\mcl{X})$.
For $Q\in\mcl{P}_n(\mcl{X})$, the corresponding type class $\mcl{T}^n(Q)$ is the set of all sequences in $\mcl{X}^n$ with type $Q$.
 The mutual information induced by the joint distribution $P$ is denoted by $I(P)$.
Finally, $N_k$ denotes the number of types in $\mcl{V}^k$. 
The analysis in this paper is largely based on the method of types \cite[Ch.~2]{csiszarkorner}.

\subsection{Preliminaries}
Following \cite{11240453}, we assume a finite collection of codeword distributions $\Q_m=\{Q_1,\ldots,Q_m\}$.
Each source type class $\mcl{T}^k(P_i)$ is assigned a distribution $Q_{\mu(i)} \in \Q_m$, and the corresponding codewords are generated from $\mcl{T}^n(Q_{\mu(i)})$ according to the proposed random construction.
Here, $\mu(i)\in[m]$ denotes the index of the codeword distribution associated with the $i$-th source type.
This setting may be viewed as partitioning the source sequences into  $m$ random-coding classes in line with \cite{6803047}. 
%
Moreover, for a source type $P_i$, we define $R_i \triangleq tH(P_i)$, since $|\mcl{T}^k(P_i)|\doteq \exp(nR_i)$ \cite[Lemma~2.3]{csiszarkorner}.

We now summarize existing results in JSCC that are relevant to this work.
The average error probability of the constant-composition ensemble satisfies \cite{csiszar1980,11240350}
\begin{align}\label{eq-random-coding-jscc}
	\bar{p}_e \,\dot{\leq} \sum_{i=1}^{N_k}
	\exp\!\left(-n\!\left[t e\!\left(\frac{R_i}{t},P_V\!\right) \!+\! E_{\mathrm{r}}(Q_{\mu(i)}, R_i)\right] \right),
\end{align}
where
\begin{align}
e(R, P_V) &\triangleq \min_{Q: H(Q) \geq R} D(Q \| P_V)\\
	\!\!\!\!E_{\mathrm{r}}(Q, R) &\triangleq \!\min_{P_{XY}:P_X=Q} D(P\|Q\times W) \!+\! \left|I_P(X;Y)\!-\!R \right|^+
\end{align}
are the source reliability function and random coding error exponent for channel coding, respectively \cite{csiszarkorner}.
This bound corresponds to the random-coding bound for JSCC.
Moreover, it can be shown that there exists a code whose error probability satisfies \cite{1056585,izs26}
\begin{align}\label{eq-exp-jscc}
	\!\!\!\!{p}_e \dot{\leq}\! \sum_{i=1}^{N_k}
	\exp\!\left(\!-n\!\left[t e\!\left(\frac{R_i}{t},P_V\!\right) \!+\! E_{\mathrm{ex}}'(Q_{\mu(i)}, \Q_m, R_i)\right] \right)
\end{align}
where
\begin{align}
	\!\!\!\!E_{\mathrm{ex}}'(Q, \Q, R) 
	\triangleq \!\!\!\!\!\! \min_{\substack{\Lambda_{X\bar{X}}\\\Lambda_X=Q, \Lambda_{\bar{X}}\in\Q\\ I_{\Lambda}(X;\bar{X})\leq R }} \!\!\!\!
	\mathbb{E}_{\Lambda}\!\left[d_W(X,\bar{X}) \right] + I(\Lambda)-R,
\end{align}
and $d_W(x,\bar{x}):=-\log\sum_y\sqrt{W(y|x)W(y|\bar{x})}$.
This bound corresponds to the expurgated bound in JSCC.

\section{Random Codebook Generation}
The construction considered here closely follows that of \cite{8656563}, and is designed to ensure that the generated codewords satisfy certain prescribed minimum-distance constraints.

Let $d(\cdot,\cdot): \mcl{X}^n\times \mcl{X}^n \rightarrow \mathbb{R}$ be a symmetric, type-dependent function that depends on $(\bs{x},\bs{x}')$ only through their joint type.
We therefore use $d(\bs{x},\bs{x}')$ and $d(\hat{P}_{\bs{x}\bs{x}'})$ interchangeably throughout this work. 
Although $d(\cdot,\cdot)$ is not necessarily a metric, it serves as a convenient distance measure in our construction.

Since the codeword distribution assigned to a source message depends on its type, we likewise allow the required minimum distance to depend on the source type.
We therefore specify a collection of real numbers $\{\Delta_i\}_{i=1}^{N_k}$, not necessarily positive, which determine the required minimum distances.
In particular, if $\bs{v} \in \mathcal{T}^k(P_i)$, then every other message $\bar{\bs{v}} \ne \bs{v}$ must satisfy $d(\bs{x}_{\bs{v}},\bs{x}_{\bar{\bs{v}}}) > \Delta_i$.
Consequently, if $\bs{v} \in \mathcal{T}^k(P_i)$ and $\bar{\bs{v}} \in \mathcal{T}^k(P_j)$, then $d(\bs{x}_{\bs{v}},\bs{x}_{\bar{\bs{v}}}) > \Delta_i \text { and } d(\bs{x}_{\bar{\bs{v}}},\bs{x}_{\bs{v}}) > \Delta_j.$
By symmetry of $d(\cdot,\cdot)$, these two conditions are equivalent to
	$d(\hat{P}_{\bs{x}_{\bs{v}}\bs{x}_{\bar{\bs{v}}}}) > \max(\Delta_i,\Delta_j)$.
We fix an arbitrary ordering of the source types, and within each type class, an arbitrary ordering of the source messages.
We denote the $\ell$-th generated codeword of source type class $\mcl{T}^k(P_i)$ by $\bs{x}^{(i)}_{\ell}$ \footnote{We use $\bs{x}$ to denote a codeword in two contexts: $\bs{x}_{\bs{v}}$ denotes the codeword corresponding to source sequence $\bs{v}$ and $\bs{x}^{(i)}_{\ell}$ denotes the $\ell$-th generated codeword of the $i$-th source type class.}.
We generate the codewords as follows:
\begin{enumerate}
	\item {Class} $\mcl{T}^k(P_1):$ 
	Choose the initial codeword $\bs{x}_1^{(1)}$ equiprobably from $\mcl{T}^n(Q_{\mu(1)})$.
	Each subsequent codeword $\bs{x}^{(1)}_{\ell}$ (for $\ell \geq 2$) is drawn equiprobably from the set
	\begin{align*}
		\!\!\!\!\left\{\bar{\bs{x}}\!\in\!\mcl{T}^n(Q_{\mu(1)})\!: d(\bs{x}^{(1)}_{\jmath}, \bar{\bs{x}})\! >\! \Delta_1 \text{ for } \jmath\in[\ell-1] \right\}.
	\end{align*}
	This continues until all $|\mcl{T}^k(P_1)|$ codewords have been generated, which we denote by $\mcl{C}_1$.
	\item {Class} $\mcl{T}^k(P_2):$ Choose the initial codeword $\bs{x}_1^{(2)}$ equiprobably from
	\begin{align*}
		\!\!\!\!\!\!\Big\{\bar{\bs{x}}\!\in\!\mcl{T}^n(Q_{\mu(2)})\!:\!d(\bs{x}, \bar{\bs{x}}) \!>\! \max(\Delta_1,\Delta_2)\, \forall \bs{x} \in \mcl{C}_1 \Big\}.
	\end{align*}
	For $\ell \geq 2$, the $\ell$-th codeword is drawn equiprobably from
	\begin{align*}
		\begin{split}
			&\!\!\!\Big\{\bar{\bs{x}}\!\in\!\mcl{T}^n(Q_{\mu(2)})\!:\!d(\bs{x}^{(2)}_{\jmath}, \bar{\bs{x}})\!>\!\Delta_2 \text{ for } \jmath\in[\ell\!-\!1],\\
			&\hspace{1.2cm}d(\bs{x}, \bar{\bs{x}}) > \max(\Delta_1,\Delta_2)\, \forall \bs{x} \in \mcl{C}_1\Big\}.
		\end{split}
	\end{align*}
	This continues until all $|\mcl{T}^k(P_2)|$ codewords have been generated, which we denote by $\mcl{C}_2$.
	\item {Class} $\mcl{T}^k(P_i):$ For $i \geq 3$, the $\ell$-th codeword is drawn equiprobably from
	\begin{align*}
			&\!\!\Big\{\bar{\bs{x}}\!\in\!\mcl{T}^n(Q_{\mu(i)})\!:\!d(\bs{x}^{(i)}_{\jmath}, \bar{\bs{x}})\!>\!\Delta_i \text{ for } \jmath\!\in[\ell\!-\!1],  \\
			&d(\bs{x}, \bar{\bs{x}}) > \max(\Delta_i,\Delta_j)\, \forall \bs{x} \in \mcl{C}_j, \text{ for } j\!\in[i\!-\!1]\Big\}\nonumber.
	\end{align*}
	Continue until all $|\mathcal{T}^k(P_i)|$ codewords have been generated; denote this set by $\mathcal{C}_i$.
\end{enumerate}
By construction, each codeword in class $i$ is at distance greater than $\Delta_i$ from all other codewords of the same class, and at distance greater than $\max(\Delta_i,\Delta_j)$ from every codeword in any class $j \neq i$.
At the end of this recursive process, the code construction has discarded no more than 
\begin{align}
	\!\!\!\!\sum_{i=1}^{N_k} \sum_{\bs{v} \in \mcl{T}^k(P_i)} \sum_{j=1}^{N_k}
	\left| \Big\{ \bar{\bs{x}} \in \mcl{T}^n(Q_{\mu(j)})\!: \!d(\bs{x}_{\bs{v}},\bar{\bs{x}}) \!\leq\! \Delta_i \Big\}\right|
\end{align}
input sequences.
In other words, this expression pessimistically counts the sequences that violate the distance constraints.
Since the distance function is type-dependent, this quantity does not depend on the specific realization of the codewords. To see this, observe
\begin{align}
	&\!\!\!\!\sum_{i=1}^{N_k} \sum_{\bs{v} \in \mcl{T}^k(P_i)} \sum_{j=1}^{N_k}
	\left| \Big\{ \bar{\bs{x}} \in \mcl{T}^n(Q_{\mu(j)}): \!d(\bs{x}_{\bs{v}},\bar{\bs{x}}) \leq \Delta_i \Big\}\right|\\
	&\hspace{0.cm}=\sum_{i=1}^{N_k} \sum_{\bs{v} \in \mcl{T}^k(P_i)} \sum_{j=1}^{N_k} \sum_{\substack{P_{X\bar{X}}: P_X=Q_{\mu(i)}\\ P_{\bar{X}}=Q_{\mu(j)}, d(P)\leq \Delta_i }} 
	\sum_{\substack{ \bar{\bs{x}} \in \mcl{T}^n(Q_{\mu(j)})\\ (\bs{x}_{\bs{v}},\bar{\bs{x}}) \in \mcl{T}^n(P) }} \!\!\! 1\\
	&\hspace{0.cm}\doteq\sum_{i=1}^{N_k} \sum_{\bs{v} \in \mcl{T}^k(P_i)} \sum_{j=1}^{N_k} \sum_{\substack{P_{X\bar{X}}: P_X=Q_{\mu(i)}\\ P_{\bar{X}}=Q_{\mu(j)}, d(P)\leq \Delta_i }}  
	\!\!\!\!\!\!\!\!\!\!\!\!\exp\left(nH_P(\bar{X}|X)\right)\\
	&\hspace{0.cm}=\sum_{(i,j)=1}^{N_k}  \sum_{\substack{P_{X\bar{X}}: P_X=Q_{\mu(i)}\\ P_{\bar{X}}=Q_{\mu(j)}, d(P)\leq \Delta_i }} \!\!\!\!\!\!\!\!\!\!
	\big|\mcl{T}^k(P_i) \big| \exp\left(nH_P(\bar{X}|X)\right).
\end{align}
Thus, we may equivalently upper bound the number of discarded sequences by
\begin{align}
	\sum_{i=1}^{N_k} \, \sum_{c=1}^{m} \, \sum_{\substack{\bar{\bs{x}}\in \mcl{T}^n(Q_{c})\\ d(\bar{\bs{x}}, \tilde{\bs{x}}_i) \leq \Delta_i }} \big|\mcl{T}^k(P_i)\big|,
\end{align}
where $\tilde{\bs{x}}_i$ is any sequence from $\mcl{T}^n(Q_{\mu(i)})$.
Then, it follows that for each class $\mcl{T}^n(Q_{c})$, $c\in[m]$, the number of sequences that violate the distance constraints is at most 
\begin{align}
	\sum_{j=1}^{N_k} \, \sum_{\substack{\bar{\bs{x}}\in \mcl{T}^n(Q_{c})\\ d(\bar{\bs{x}}, \tilde{\bs{x}}_j) \leq \Delta_j }} \big|\mcl{T}^k(P_j)\big|.
\end{align}
To ensure that the code construction produces the required number of valid codewords, we assume that for each class $c \in [m]$, non-negative values $\{\delta_c\}_{c=1}^{m}$ are chosen such that
\begin{align}\label{eq-cond}
	\sum_{j=1}^{N_k} \, \sum_{\substack{\bar{\bs{x}}\in \mcl{T}^n(Q_{c})\\ d(\bar{\bs{x}}, \tilde{\bs{x}}_j) \leq \Delta_j }} \big|\mcl{T}^k(P_j)\big|
	&\leq \zeta_n \left| \mcl{T}^n(Q_{c}) \right| e^{-n\delta_c},
\end{align}
where $\zeta_n:=N_k(n+1)^{|\mcl{X}|^2+|\mcl{X}|}$.
For simplicity, and without loss of generality, we take $\delta_c=\delta$ for all $c\in[m]$ and some $\delta>0$.
Condition \eqref{eq-cond} guarantees that, for each source type class $\mcl{T}^k(P_i)$
the set from which the codewords are drawn equiprobably has cardinality at least
\begin{align}
	\begin{split}\label{eq:card-lb}
		&\left| \mcl{T}^n(Q_{\mu(i)}) \right| - \zeta_n \left| \mcl{T}^n(Q_{\mu(i)}) \right| e^{-n\delta}\\
		&\hspace{2cm}\geq \left(1-\zeta_n\,e^{-n\delta}\right) \left| \mcl{T}^n(Q_{\mu(i)}) \right| 
	\end{split}
\end{align}
at every step of the construction.
Therefore, whenever \eqref{eq-cond} holds, the recursive code construction produces the required number of codewords $|\mcl{V}|^k$.

\begin{lemma}\label{lem:marg}
Let $\bs{v} \in \mcl{T}^k(P_i)$. Then the marginal distribution of the corresponding
codeword $\bs{X}_{\bs{v}}$ is uniform over $\mcl{T}^n(Q_{\mu(i)})$, 
\begin{align}
\mathbb{P}\!\left[\bs{X}_{\bs{v}}=\bs{x}\right]
=
\frac{1}{\left|\mcl{T}^n(Q_{\mu(i)})\right|}
\quad \text{for } \bs{x} \in \mcl{T}^n(Q_{\mu(i)}),
\end{align}
and zero otherwise.
\end{lemma}

\begin{lemma}\label{lem:join}
	Let $\bs{v} \in \mcl{T}^k(P_i)$ and $\tilde{\bs{v}} \in \mcl{T}^k(P_j)$, and let $\bs{x} \in \mcl{T}^n(Q_{\mu(i)})$ and $\tilde{\bs{x}} \in \mcl{T}^n(Q_{\mu(j)})$
	denote their corresponding codewords under the proposed codebook construction.
	If $d(\bs{x}, \tilde{\bs{x}}) > \max(\Delta_i,\Delta_j)$, then
	\begin{align}
		\begin{split}
			&\mathbb{P}\left[\left(\bs{X}_{\bs{v}}, \bs{X}_{\tilde{\bs{v}}} \right)=\left(\bs{x},\tilde{\bs{x}}\right) \right] \\
			&~~~~~~  \leq\frac{1}{ \left|\mcl{T}^n(Q_{\mu(i)}) \right|\cdot \left|\mcl{T}^n(Q_{\mu(j)}) \right| \cdot\left(1-\zeta_n e^{-n\delta}\right)^2}
		\end{split}
	\end{align}
	while $\mathbb{P}\left[\left(\bs{X}_{\bs{v}}, \bs{X}_{\tilde{\bs{v}}} \right)=\left(\bs{x},\tilde{\bs{x}}\right) \right]\!\!=\!\!0$ whenever $d(\bs{x}, \tilde{\bs{x}}) \leq \max(\Delta_i,\Delta_j)$.
\end{lemma} 
The proof of Lemma~\ref{lem:marg} follows identically to that of \cite[Lemma~4]{8656563}.
The proof of Lemma~\ref{lem:join}, along with the remaining proofs of the paper, is provided in Section~\ref{sec:proof}.

\section{Achievable Error Exponents}
Consider a type-dependent metric, meaning that it depends on the source and channel sequences only through their types.
Specifically, for $\bs{v}\in\mathcal{T}^k(P_i)$, the metric can be expressed as $q(\bs{v},\bs{x},\bs{y}) = q\big(i,\hat{P}_{\bs{x}\bs{y}}\big)$.

\begin{theorem}\label{thm:main}
	Consider the code construction described with parameters $\big(d,\{\Delta_i\}_{i=1}^{N_k},\delta\big)$  satisfying \eqref{eq-cond}, together with a type-dependent decoding metric $q$.
	 The ensemble average error probability satisfies
	\begin{align}
		\bar{p}_e \,\dot{\leq}\,
		\sum_{i=1}^{N_k} \sum_{j=1}^{N_k}
		\exp\!\left(-n E_{\mathrm{rgv}}^{(ij)}(P_V, W)\right),
		\label{eq:rgv_pe}
	\end{align}
	where
	\begin{align}\label{eq:rgv-expression}
			\!\!\!\!E_{\mathrm{rgv}}^{(ij)}(P_V, W) \!&\triangleq te\!\left(\frac{R_i}{t},P_V\!\right)\!+ \!\!\min_{P \in \Gamma_{ij}}\!\! \Big\{D\!\left(P_{XY}\middle\| Q_{\mu(i)}\!\times\! W\right) \nonumber\\
			&\qquad\qquad + \big| I_P(\bar{X}; X,Y) - R_j \big|^{+}\Big\},
	\end{align}
	and
	\begin{align}\label{eq:gamma}
		&\Gamma_{ij} \triangleq \!\bigg\{P_{X\bar{X}Y}: \!P_X=Q_{\mu(i)}, P_{\bar{X}}=Q_{\mu(j)},\nonumber\\ 
		&\hspace{0.1cm}d(P_{X\bar{X}}) \geq \max(\Delta_i,\Delta_j), q(i, P_{XY})\leq q(j, P_{\bar{X}Y})  \bigg\}.
	\end{align}
\end{theorem}

We next show that with an appropriate choice of the distance function and the minimum-distance parameters, the ensemble average error probability in \eqref{eq:rgv_pe} achieves both the random-coding and expurgated error exponents.
To this end, following the choices in \cite[Sec.~VI]{8656563}, we set $d(P) = -I(P)$, and $\Delta_i = -(R_i + \delta)$. 
We will show that these choices satisfy the condition in \eqref{eq-cond}.
Observe that, for each $c\in[m]$,
\begin{align}
	\begin{split}
		&\sum_{j=1}^{N_k} \, \sum_{\substack{\bar{\bs{x}}\in \mcl{T}^n(Q_{c})\\ d(\bar{\bs{x}}, \tilde{\bs{x}}_j) \leq \Delta_j }} \big|\mcl{T}^k(P_j)\big|\\
		&\hspace{0.55cm}\leq \sum_{j=1}^{N_k} \, \sum_{\substack{ P_{X\bar{X}}: P_X=Q_{c}\\P_{\bar{X}}=Q_{\mu(j)}\\-I(P)\leq -(R_j+\delta)  }}  
		\sum_{\substack{{\bs{x}}\in \mcl{T}^n(Q_{c})\\ ({\bs{x}}, \tilde{\bs{x}}_j)\in \mcl{T}^n(P) \ }} e^{nR_j}\\
	\end{split}\\
	&\hspace{0.55cm}\leq \sum_{j=1}^{N_k} \, \sum_{\substack{ P_{X\bar{X}}: P_X=Q_{c}\\P_{\bar{X}}=Q_{\mu(j)}\\-I(P)\leq -(R_j+\delta)  }}  \!\!\!\!\!\!
	\frac{e^{nH_P(X|\bar{X})} e^{nR_j}}{|\mcl{T}^n(Q_{c})|} |\mcl{T}^n(Q_{c})|.
\end{align}
Recall from \cite[Lemma~2.3]{csiszarkorner} that for any type $Q$, we have $(n+1)^{-|\mcl{X}|} e^{nH(Q)} \leq |\mcl{T}^n(Q)|$. 
Therefore,
\begin{align}
	\frac{e^{n\left(H_P(X|\bar{X})+R_j\right)}}{|\mcl{T}^n(Q_{c})|} \leq (n+1)^{|\mcl{X}|} e^{-n\left(I_P(X;\bar{X})-R_j\right)}.
\end{align}
Moreover, by the minimum-distance condition, any admissible distribution $P$ satisfies $I(P)-R_j \geq \delta$. Thus,
\begin{align}
	\frac{e^{n\left(H_P(X|\bar{X})+R_j\right)}}{|\mcl{T}^n(Q_{c})|} \leq (n+1)^{|\mcl{X}|} e^{-n\delta}.
\end{align}
Substituting this back, we obtain
\begin{align}
	\begin{split}
		&\sum_{j=1}^{N_k} \, \sum_{\substack{\bar{\bs{x}}\in \mcl{T}^n(Q_{c})\\ d(\bar{\bs{x}}, \tilde{\bs{x}}_j) \leq \Delta_j }} \big|\mcl{T}^k(P_j)\big|\\
		&\hspace{1.3cm}\leq \sum_{j=1}^{N_k} \, \sum_{\substack{ P_{X\bar{X}}: P_X=Q_{c}\\P_{\bar{X}}=Q_{\mu(j)}\\-I(P)\leq -(R_j+\delta)  }}\!\!\!\!\!\!\!\! (n+1)^{|\mcl{X}|} e^{-n\delta}
	\end{split}\\
	&\hspace{1.3cm}\leq N_k (n+1)^{|\mcl{X}|^2+|\mcl{X}|} e^{-n\delta},
\end{align}
thereby satisfying \eqref{eq-cond}. Therefore, the minimum-distance condition reduces to $I(P) < \min(R_i,R_j)+\delta$.
With these choices, the ensemble is universal in the sense that it does not depend on either the source or the channel distribution, as in \cite{8656563}.
Moreover, the resulting ensemble achieves the random-coding and expurgated exponents simultaneously.

\subsection{Achieving the Random Coding Error Exponent}

To attain the random-coding exponent, we choose the decoding metric to be the universal generalized maximum mutual information introduced by Csisz{\'a}r \cite{csiszar1980}, namely,
\begin{align}
	q(i, P) = I(P)-R_i.
\end{align}
We  relax the constraint set $\Gamma_{ij}$ in \eqref{eq:gamma} by dropping the minimum-distance requirement, which can only weaken the bound.
In addition, we use the inequality $I_P(\bar{X};X,Y) \geq I_P(\bar{X};Y)$ to further lower bound the exponent.
This yields,
\begin{align}
		\!\!\!E_{\mathrm{rgv}}^{(ij)}(P_V,\!W) &\!\geq\! te\!\left(\frac{R_i}{t},P_V\!\right)\!+ \min_{P} \Big\{D\!\left(P_{XY}\middle\| Q_{\mu(i)}\!\times\! W\right)\nonumber\\
		&\qquad\qquad + \big| I_P(\bar{X}; Y) - R_j \big|^{+}\Big\},
\end{align}
where the minimization is over distributions $P$ satisfying
$P_X = Q_{\mu(i)}$, $P_{\bar{X}} = Q_{\mu(j)}$, and
$I_P(X;Y) - R_i \leq I_P(\bar{X};Y) - R_j$.
Hence, for any admissible $P$,
\begin{align}
	\big| I_P(\bar{X}; Y) - R_j \big|^{+} \geq \big| I_P({X}; Y) - R_i \big|^{+}.
\end{align}
Therefore, for all $j \in [N_k]$, 
\begin{align}
	\begin{split}
		&E_{\mathrm{rgv}}^{(ij)}(P_V,W) \!\geq  te\!\left(\frac{R_i}{t},P_V\!\right) + E_{\mathrm{r}}(Q_{\mu(i)}, R_i).
	\end{split}
\end{align}
Consequently,
\begin{align}
	\bar{p}_e \,\dot{\leq} \sum_{i=1}^{N_k}
	\exp\!\left(-n\!\left[te\!\left(\frac{R_i}{t},P_V\!\right) \!+\! E_{\mathrm{r}}(Q_{\mu(i)}, R_i)\right] \right)
\end{align}
which matches the JSCC random coding bound (see \eqref{eq-random-coding-jscc}).

\subsection{Achieving the Expurgated Error Exponent}
To attain the expurgated error exponent, we adopt the decoding metric proposed in \cite{1056585}. In particular, we set
\begin{align}
	q(i, P) = \mathbb{E}_P\big[\log W(Y|X)\big]-2R_i.
\end{align}
We lower bound the expression in \eqref{eq:rgv-expression} by removing the $|\cdot|^{+}$ operator.
For any admissible distribution $P \in \Gamma_{ij}$, 
\begin{align}
	\begin{split}
		&D(P_{XY}\|Q_{\mu(i)}\times W) + I_P(\bar{X};X,Y)\\
		&\hspace{1.9cm}= D(P \| P_{X\bar{X}}\times W) + I_P(X;\bar{X}).
	\end{split}
\end{align}
Using this identity, we obtain the following lower bound,
\begin{align}
	&\!\!\!\!E_{\mathrm{rgv}}^{(ij)}(P_V,W) \geq te\left(\frac{R_i}{t},P_V\right) +\!\!\!\!\!\!\! \min_{\substack{\Lambda_{X\bar{X}}: \Lambda_X=Q_{\mu(i)}\\\Lambda_{\bar{X}}=Q_{\mu(j)}\\I(\Lambda)\leq \min(R_i,R_j)+\delta}}
		\!\!\!\!\!\!\! I_{\Lambda}(X;\bar{X}) \nonumber\\
		&~~~~~~~~~~~~+\!\!\!\!\!\!\!\!\! \min_{\substack{P: P_{X\bar{X}}=\Lambda\\ \mathbb{E}_P\big[\log \frac{W(Y|X)}{W(Y|\bar{X})}\big]\leq 2(R_i-R_j) }}\!\!\!\!\!\!\!\!\!\!\!\!
		D(P\|\Lambda \times W) - R_j.
\end{align}
We now evaluate the inner minimization. By Lagrange duality,
\begin{align}
	\begin{split}
		&\min_{\substack{P: P_{X\bar{X}}=\Lambda\\\mathbb{E}_P\!\log \big[\frac{W(Y|X)}{W(Y|\bar{X})}\big]\leq 2(R_i-R_j) }}\!\!\!\!\!\!\!\!\!\!\!\! D(P\|\Lambda \times W) - R_j\\
		&=\min_{\substack{P: P_{X\bar{X}}=\Lambda}} \, \sup_{s \geq 0}\, \mathbb{E}_P \bigg[\log \frac{P(Y|X,\bar{X})}{W(Y|X)^{1-s}W(Y|\bar{X})^s} \bigg]\\
		&~~~~~~~~~~~~~ - R_j + 2s(R_j - R_i).
	\end{split}
\end{align}
Choosing $s=\tfrac{1}{2}$, which may be suboptimal, and applying the log-sum inequality yields
\begin{align}
	&\min_{\substack{P: P_{X\bar{X}}=\Lambda\\\mathbb{E}_P \big[\log \frac{W(Y|X)}{W(Y|\bar{X})}\big]\leq 2(R_i-R_j) }}\!\!\!\!\!\!\!\!\!\!\!\! D(P\|\Lambda \times W) - R_j\\
		&\hspace{0.45cm}\geq -\sum_{(x,\bar{x})} \Lambda(x,\bar{x})\log \sum_y \sqrt{W(y|x)W(y|\bar{x})} -R_i. \nonumber \label{eq:logsum1}
\end{align}
Therefore,
\begin{align}
	E_{\mathrm{rgv}}^{(ij)}(t&,P_V) \geq te\!\left(\frac{R_i}{t},P_V\right) \\ &  +\!\!\! \min_{\substack{\Lambda_{X\bar{X}}: \Lambda_X=Q_{\mu(i)}\\\Lambda_{\bar{X}}=Q_{\mu(j)}\\I(\Lambda)\leq \min(R_i,R_j)+\delta}}
		\!\!\!\!\! I(\Lambda)\!+
	\mathbb{E}_{\Lambda}[d_W(X,\bar{X})] \!-\!R_i. \nonumber
\end{align}
We can further relax the bound by replacing the constraint $I(\Lambda)\le \min(R_i,R_j)+\delta$ with $I(\Lambda)\le R_i+\delta$, and by relaxing the condition $\Lambda_{\bar{X}}=Q_{\mu(j)}$ to $\Lambda_{\bar{X}}\in\mathcal{Q}_m$.
The bound becomes independent of $j$.
For all $j\in[N_k]$, letting $\delta\to0$ yields
\begin{align}
	\begin{split}
		&\!\!\!\!E_{\mathrm{rgv}}^{(ij)}(P_V,W) \geq te\!\left(\frac{R_i}{t},P_V\right) + E_{\mathrm{ex}}'(Q_{\mu(i)}, \Q_m, R_i).
	\end{split}
\end{align}
Therefore,
\begin{align}
	\!\!\!\!\!{p}_e \,\dot{\leq} \sum_{i=1}^{N_k}\!
	\exp\!\left(\!-n\!\left[te\!\left(\frac{R_i}{t},P_V\!\right) \!+\! E_{\mathrm{ex}}'(Q_{\mu(i)}, \Q_m, R_i)\right] \right)
\end{align}
which matches the expurgated bound in \eqref{eq-exp-jscc}.

	The above recursive construction naturally applies to single-class coding, in which all codewords are assigned the same distribution, i.e., $Q_c = Q$ for all $c\in[m]$.
	In that case, the corresponding single-class results are recovered; see \cite{11240453}.


\section{Proofs}\label{sec:proof}
\emph{Proof of Lemma \ref{lem:join}.}
	The proof closely follows that of \cite[Lemma~2]{8656563}.
	Without loss of generality, assume $i<j$.
	Let $\bs{v}$ and $\tilde{\bs{v}}$ be the $\ell_i$-th and $\ell_j$-th sequences in $\mcl{T}^k(P_i)$ and $\mcl{T}^k(P_j)$, respectively.
	By construction, the codeword $\bs{X}_{\bs{v}}$ is drawn equiprobably from the set $\mcl{T}^n(Q_{\mu(i)}, i, \ell_{i}) \subseteq \mcl{T}^n(Q_{\mu(i)})$,
	consisting of all sequences that satisfy the imposed distance constraints with respect to all previously generated codewords. Consequently, for any admissible $\bs{x}$,
	\begin{align}\label{eq:joint-up-1}
		\mathbb{P}\!\left[\bs{X}_{\bs{v}}=\bs{x}\,\big|\,\text{past codewords}\right]
		\le \frac{1}{\left|\mcl{T}^n(Q_{\mu(i)},i,\ell_i)\right|}.
	\end{align}
	An analogous argument applies to $\bs{X}_{\tilde{\bs{v}}}$.
	Expanding the joint probability  yields
	\begin{align}\label{eq:joint-lemma}
		&\mathbb{P}\left[\left(\bs{X}_{\bs{v}}, \bs{X}_{\tilde{\bs{v}}} \right)=\left(\bs{x},\tilde{\bs{x}}\right) \right]\nonumber\\&=
		\sum_{(\text{before } \bs{v})} \sum_{(\text{after } \bs{v} \text{ before } \tilde{\bs{v}})}\mathbb{P}\Big[\text{codewords before } \bs{v} \Big]\nonumber\\
		&\hspace{0.55cm}\times\mathbb{P}\Big[\bs{X}_{\bs{v}}=\bs{x} \,\big| \,\text{codewords before } \bs{v} \Big]\\
		&\hspace{0.55cm}\times\mathbb{P}\Big[\text{codewords after } \bs{v}\text{ before }\tilde{\bs{v}} \,\big| \,\text{past codewords} \Big]\nonumber\\
		&\hspace{0.55cm}\times\mathbb{P}\Big[\bs{X}_{\tilde{\bs{v}}}=\tilde{\bs{x}} \,\big| \,\text{codewords before } \tilde{\bs{v}} \Big]\nonumber.
	\end{align}
	Using \eqref{eq:joint-up-1} to upper bound the second and
	fourth probability terms in \eqref{eq:joint-lemma}, and noting that the sums over
	the remaining probability terms evaluate to one, we obtain
	\begin{align}
		\begin{split}
			&\mathbb{P}\left[\left(\bs{X}_{\bs{v}}, \bs{X}_{\tilde{\bs{v}}} \right)\!=\!\left(\bs{x},\tilde{\bs{x}}\right) \right]\\ &\hspace{1.6cm}\leq
			\frac{1}{\left|\mcl{T}^n(Q_{\mu(i)}, i, \ell_{i}) \right|}\frac{1}{\left|\mcl{T}^n(Q_{\mu(j)}, j, \ell_{j}) \right|}.
		\end{split}
	\end{align}
	Lower bounding each cardinality by \eqref{eq:card-lb} yields the result.
\emph{Proof of Theorem \ref{thm:main}.}
	We begin with the extended RCU bound, given in \cite[Eq.~(26)]{6803047},
	\begin{align}
		\bar{p}_e
		\,\dot{\leq} \, \sum_{i=1}^{N_k} \bar{\varepsilon}_i,
	\end{align}
	where
	\begin{align}
	\begin{split}
		&\bar{\varepsilon}_i =\!\!\!\! \sum_{\bs{v}\in \mcl{T}^k(P_i)}  P^k(\bs{v}) \, \sum_{(\bs{x},\bs{y})} \mathbb{P}\left[\bs{X}_{\bs{v}}=\bs{x}\right] W^n(\bs{y}|\bs{x})\\
		&\,\,\min\Bigg\{1, \sum_{j=1}^{N_k}\sum_{\bar{\bs{v}} \in \mcl{T}^k(P_j)}\!\!\!\!\!
		\mathbb{P}\left[{q(j, \hat{P}_{\bs{X}_{\bar{\bs{v}}}\bs{y}})}\geq q(i, \hat{P}_{\bs{x}\bs{y}}) \right]\Bigg\}.
	\end{split}
	\end{align}
	For a given $(\bs{v}, \bar{\bs{v}}, \bs{x}, \bs{y})$, and using Lemmas~\ref{lem:marg} and~\ref{lem:join}, the pairwise error probability can be written as
	\begin{align}
		&\mathbb{P}\left[q(P_j, \hat{P}_{\bs{X}_{\bar{\bs{v}}}\bs{y}})\geq q(P_i, \hat{P}_{\bs{x}\bs{y}}) \right] \nonumber \\
		&= \sum_{\substack{\bar{\bs{x}}\in \mcl{T}^n(Q_{\mu(j)})\\ d(\bs{x},\bar{\bs{x}})>\max(\Delta_i,\Delta_j)\\q(j, \hat{P}_{\bar{\bs{x}}\bs{y}})\geq q(i, \hat{P}_{\bs{x}\bs{y}})}}
		\mathbb{P}\left[\bs{X}_{\bar{\bs{v}}} = \bar{\bs{x}}\, \big| \, \bs{X}_{\bs{v}}=\bs{x},\bs{Y}=\bs{y} \right]\\
		&= \sum_{\substack{\bar{\bs{x}}\in \mcl{T}^n(Q_{\mu(j)})\\ d(\bs{x},\bar{\bs{x}})>\max(\Delta_i,\Delta_j)\\q(j, \hat{P}_{\bar{\bs{x}}\bs{y}})\geq q(i, \hat{P}_{\bs{x}\bs{y}})}}
		\frac{\mathbb{P}\left[\bs{X}_{\bar{\bs{v}}} = \bar{\bs{x}} ,  \bs{X}_{\bs{v}}=\bs{x}\right]}{\mathbb{P}\left[\bs{X}_{\bs{v}}=\bs{x} \right]}\\
		&\leq\, \frac{1}{\left(1-\xi_ne^{-n\delta}\right)^2} \!\!\!
		  \sum_{\substack{\bar{\bs{x}}\in \mcl{T}^n(Q_{\mu(j)})\\ d(\bs{x},\bar{\bs{x}})>\max(\Delta_i,\Delta_j)\\q(j, \hat{P}_{\bar{\bs{x}}\bs{y}})\geq q(i, \hat{P}_{\bs{x}\bs{y}})}}
		\frac{1}{|\mcl{T}^n(Q_{\mu(j)})|}. \label{eq-w-pre}
	\end{align}

	Observe that the bound does not depend on the particular $\bar{\bs{v}}$. Moreover, the factor $(1-\xi_n e^{-n\delta})$ can be made arbitrarily close to one for sufficiently large $n$. Thus, we obtain
	\begin{align}
		\begin{split}
			&\sum_{\bar{\bs{v}} \in \mcl{T}^k(P_j)}\mathbb{P}\left[q(j, \hat{P}_{\bs{X}_{\bar{\bs{v}}}\bs{y}})\geq q(i, \hat{P}_{\bs{x}\bs{y}}) \right]  \\
			&\hspace{2cm}\dot{\leq}\,\sum_{\substack{\bar{\bs{x}}\in \mcl{T}^n(Q_{\mu(j)})\\ 
			d(\bs{x},\bar{\bs{x}})>\max(\Delta_i,\Delta_j)\\q(j, \hat{P}_{\bar{\bs{x}}\bs{y}})\geq q(i, \hat{P}_{\bs{x}\bs{y}})}}
			\frac{e^{nR_j}}{|\mcl{T}^n(Q_{\mu(j)})|}.
		\end{split}
	\end{align}
	Using the method of types, we have
	\begin{align}
		&\sum_{\substack{\bar{\bs{x}}\in \mcl{T}^n(Q_{\mu(j)})\\  d(\bs{x},\bar{\bs{x}})>\max(\Delta_i,\Delta_j)\\q(j, \hat{P}_{\bar{\bs{x}}\bs{y}})\geq q(i, \hat{P}_{\bs{x}\bs{y}})}}
		\frac{e^{nR_j}}{|\mcl{T}^n(Q_{\mu(j)})|}\\
		&=\!\!\!\!\!\!\sum_{\substack{V_{X\bar{X}Y}: V_{XY}=\hat{P}_{\bs{xy}}, V_{\bar{X}}=Q_{\mu(j)}\\ d(V_{X\bar{X}})>\max(\Delta_i,\Delta_j)\\ q(j, V_{\bar{X}Y})\geq q(i, {V}_{XY})}}
		\sum_{\substack{\bar{\bs{x}} \in \mcl{T}^n(Q_{\mu(j)})\\(\bs{x},\bar{\bs{x}},\bs{y})\in \mcl{T}^n(V)}} \frac{e^{nR_j}}{|\mcl{T}^n(Q_{\mu(j)})|}\\
		&\doteq\!\!\!\!\!\!\sum_{\substack{V_{X\bar{X}Y}: V_{XY}=\hat{P}_{\bs{xy}}, V_{\bar{X}}=Q_{\mu(j)}\\ d(V_{X\bar{X}})>\max(\Delta_i,\Delta_j)\\ q(j, V_{\bar{X}Y})\geq q(i, {V}_{XY})}}
		\frac{e^{nH_V(\bar{X}|X,Y)} e^{nR_j}}{e^{nH_V(\bar{X})}} \label{eq-mot-2} \\
		&\doteq\!\!\!\!\!\!\!\!\!\max_{\substack{V_{X\bar{X}Y}: V_{XY}=\hat{P}_{\bs{xy}}, V_{\bar{X}}=Q_{\mu(j)}\\ d(V_{X\bar{X}})>\max(\Delta_i,\Delta_j)\\ q(j, V_{\bar{X}Y})\geq q(i, {V}_{XY})}}
		\!\!\!\!\!\!\!\!\!\!\!\exp\left(-n\left[I_V(\bar{X} ; X, Y) - R_j \right] \right). \label{eq-mot-3}
	\end{align}
	Similarly, $\mcl{T}^n(Q_{\mu(j)}) \doteq \exp\!\left(n H_V(\bar{X})\right)$.
	Furthermore, observe that
	 for given $(\bs{x},\bs{y})$ where
	$\bs{x}\in \mcl{T}^n(Q_{\mu(i)})$, we can evaluate the expectation as follows
	\begin{align}
		\begin{split}
			&\sum_{(\bs{x},\bs{y})} \mathbb{P}\left[\bs{X}_{\bs{v}}=\bs{x}\right] W^n(\bs{y}|\bs{x})\\
			&\hspace{0.45cm}\doteq \max_{ \substack{ P_{XY}: P_X=Q_{\mu(i)} }} \exp\left(-n D(P \,\|\, Q_{\mu(i)}\! \times W) \right).
		\end{split}
	\end{align}
	The proof is completed by observing, following \cite[Eqs. (45)–(48)]{6283524}, that 
	\begin{align}
		\sum_{\bs{v}\in \mcl{T}^k(P_i)}  P^k(\bs{v}) \leq \exp\left(\!-k e\!\left(\frac{R_i}{t},P_V \right) \right).
	\end{align}
	Therefore, overall, 
	\begin{align}
		&\!\!\!\bar{\varepsilon}_i
			\,\dot{\leq}\, \sum_{j=1}^{N_k} \exp\!\bigg(\!-n\bigg[
			te\!\left(\!\frac{R_i}{t},P_V\!\right)\!+ \!\min_{P \in \Gamma_{ij}}\! \Big\{\!D\!\left(P_{XY}\middle\| Q_{\mu(i)}\!\times\! W\right) \nonumber \\
			& \hspace{2cm} + \big| I_P(\bar{X}; X,Y) - R_j \big|^{+}\Big\} \bigg] \bigg).
	\end{align}


{\footnotesize
\bibliographystyle{IEEEtran}
\bibliography{refs}
}

\end{document}